\documentclass[aps,prb,twocolumn]{revtex4}
\usepackage{amssymb}
\usepackage{graphicx}

\begin{document}

\title{Quantum Griffiths effects in itinerant Heisenberg magnets}
\author{Thomas Vojta}
\affiliation{Department of Physics, University of Missouri-Rolla, Rolla, MO 65409}
\author{J\"org Schmalian}
\affiliation{Department of Physics and Astronomy and Ames Laboratory, Iowa State
University, Ames, IA 50011}
\date{\today}

\begin{abstract}
We study the influence of quenched disorder on quantum phase transitions in
itinerant magnets with Heisenberg spin symmetry, paying particular attention
to rare disorder fluctuations. In contrast to the Ising case where the
overdamping suppresses the tunneling of the rare regions, the Heisenberg
system displays strong power-law quantum Griffiths singularities in the
vicinity of the quantum critical point. We discuss these phenomena based on
general scaling arguments, and we illustrate them by an explicit calculation
for $O(N)$ spin symmetry in the large-$N$ limit. We also discuss broad
implications for the classification of quantum phase transitions in the
presence of quenched disorder.
\end{abstract}

\maketitle




\section{Introduction}

\label{sec:introduction}

At low temperatures, strongly correlated materials can display a surprising
sensitivity to small amounts of imperfections and disorder. This effect is
particularly pronounced close to a quantum phase transition, where large
fluctuations in space and time become fundamentally connected. Thermodynamic
and response properties of a material are then affected much more
dramatically than close to a classical phase transition, permitting exotic
phenomena like infinite-randomness critical points with activated rather
than power-law dynamical scaling \cite%
{McCoy69,dsf9295,YoungRieger96,Pich98,Motrunich00,ThillHuse95,gbh96,RiegerYoung96}%
, smeared transitions \cite{smearing_prl}, or non-universal exponents at
certain impurity quantum phase transitions \cite{kondo}.

One interesting aspect of phase transitions in disordered systems are the
Griffiths effects \cite{Griffiths}. They are caused by large spatial regions
that are devoid of impurities and can show local order even if the bulk
system is in the disordered phase. The fluctuations of these regions are
very slow because they require changing the order parameter in a large
volume. Griffiths \cite{Griffiths} showed that this leads to a singular free
energy in a whole parameter region in the vicinity of the critical point
which is now known as the Griffiths phase. In generic classical systems, the
contribution of the rare regions to thermodynamic observables is very weak
since the singularity in the free energy is only an essential one \cite%
{Griffiths,Bray89,Randeria}. The consequences for the dynamics are more
severe with the rare regions dominating the long-time behavior \cite%
{Randeria,Bray87,Dhar88}.

Due to the perfect disorder correlations in (imaginary) time, Griffiths
phenomena at zero temperature quantum phase transitions are enhanced
compared to their classical counterparts. In random quantum Ising systems
\cite{McCoy69,dsf9295,YoungRieger96,Pich98,Motrunich00} and quantum Ising
spin glasses \cite{ThillHuse95,gbh96,RiegerYoung96}, thermodynamic
quantities display power-law singularities with continuously varying
exponents in the Griffiths phase, with the average susceptibility actually
diverging inside this region.

The systems in which these quantum Griffiths phenomena have been shown
unambiguously all have undamped dynamics (a dynamical exponent $z=1$ in the
corresponding clean system). However, many systems of experimental
importance involve magnetic \cite%
{Lohneysen96,Mackenzie01,Lonzarich99,Maple98} or superconducting \cite%
{Rogachev03} degrees of freedom coupled to conduction electrons which leads
to overdamped dynamics characterized by a clean dynamical exponent $z>1 $.
Studying the effects of rare regions in this case is therefore an important
issue. In recent years, there has been an intense debate on the theory of
quantum Griffiths effects in itinerant \emph{Ising} magnets. It has been
suggested\cite{CastroNetoJones} that overdamped systems show quantum
Griffiths phenomena similar to that of undamped systems. However, recently
it has been shown\cite{MillisMorrSchmalian,smearing_prl} that the
overdamping prevents the rare regions from tunneling leading to static rare
regions displaying superpara\-magnetic rather than quantum Griffiths
behavior, at least for sufficiently low temperatures. In Ref.\cite%
{MillisMorrSchmalian} it was also pointed out that different behavior is
expected for systems with continuous spin symmetry.

In this paper, we examine quantum Griffiths effects in itinerant \emph{%
Heisenberg} magnets. Our results can be summarized as follows: In contrast
to the Ising case, itinerant magnets with Heisenberg symmetry (in general, $%
O\left( N\right) $ symmetry with $N>1$) do display power-law quantum
Griffiths singularities. The locally ordered rare regions are not static but
retain their quantum dynamics. Their low-energy density of states follows a
power law, $\rho (\epsilon )\sim \epsilon ^{d/z^{\prime }-1}$ where $d$ is
the space dimensionality and $z^{\prime }$ is a continuously varying
dynamical exponent. This leads to power-law dependencies of several
observables on the temperature $T$, including the specific heat, $C\sim
T^{d/z^{\prime }}$, and the magnetic susceptibility, $\chi \sim
T^{d/z^{\prime }-1}$. Our results are not limited to Heisenberg magnets,
they generally apply to $O\left( N\right)$ order parameters with $N>1$
including the superconductor-metal transition \cite{Sachdev04} in thin
nano-wires \cite{Rogachev03}.

The paper is organized as follows: In section \ref{sec:scaling}, we derive
quantum Griffiths effects from general scaling arguments based on the
observation that a rare region in an itinerant Heisenberg magnet is at its
lower critical dimension. These arguments suggest a general classification
of disordered quantum phase transitions in terms of the dimensionality of
the rare regions. In section \ref{sec:large-N}, we then present an explicit
calculation for $O(N)$ spin symmetry in the large-$N$ limit. We conclude in
section \ref{sec:conclusions} by discussing the importance of the spin
symmetry, the relation to Kondo physics as well as possible experimental
realizations of our predictions.


\section{Quantum Griffiths effects from scaling arguments}

\label{sec:scaling}

Our starting point is a quantum Landau-Ginzburg-Wilson free energy
functional for an $N$-component ($N>1$) order parameter field $\phi =(\phi
_{1},\ldots ,\phi _{N})$. For definiteness, we consider the itinerant
antiferromagnetic transition. The action of the clean system reads \cite%
{BelitzKirkpatrick96,Hertz76,Millis93}
\begin{equation}
S=\int dx\,dy\ \phi (x)\,\Gamma (x,y)\,\phi (y)+\frac{u}{2N}\int dx\ \phi
^{4}(x)~.  \label{eq:action}
\end{equation}%
Here, $x\equiv (\mathbf{x},\tau )$ comprises position $\mathbf{x}$ and
imaginary time $\tau $, and $\int dx\equiv \int d\mathbf{x}%
\int_{0}^{1/T}d\tau $. The imaginary time direction which characterizes the
quantum dynamics formally appears like an additional spatial dimension of a
classical system. The perfect disorder correlations in this imaginary time
direction are ultimately responsible for the enhancement of the Griffiths
effects at zero temperature. $\Gamma (x,y)$ is the bare inverse propagator
(bare two-point vertex), whose Fourier transform is
\begin{equation}
\Gamma (\mathbf{q},\omega _{n})=(r_{0}+\mathbf{q}^{2}+\gamma |\omega
_{n}|^{2/z})
\end{equation}
and $r_{0}$ is the bare energy gap, i.e., the bare distance from the clean
critical point.

We are interested in the case of overdamped spin dynamics ($z=2$) with $%
\gamma \simeq \left( J\rho _{F}\right) ^{2}/\left( E_{F}a_{0}^{2}\right) $
where $J$ is the coupling constant between spin degrees of freedom and
conduction electrons, with DOS, $\rho _{F}$, and Fermi energy, $E_{F}$,
respectively. $a_{0}$ is the lattice constant. In order to demonstrate the
special behavior of Heisenberg systems with $z=2$, where overdamping is
caused by the particle-hole continuum of itinerant electrons, we frequently
discuss variable $z$ and compare the behavior with ballistic spin systems
with $z=1$. We will use a system of units with $\gamma =1$. The clean system
undergoes the quantum phase transition when the renormalized gap $r$
vanishes. To introduce quenched disorder, we dilute the system with
nonmagnetic impurities of spatial density $p$, i.e., we add a random
potential, $\delta r(\mathbf{x})=\sum_{i}V[\mathbf{x}-\mathbf{x}(i)]$, to $%
r_{0}$. Here, $\mathbf{x}(i)$ are the the positions of the impurities, and $%
V(\mathbf{x})$ is a positive short-ranged impurity potential.

We first present the general scaling arguments leading to quantum Griffiths
behavior in this system. Despite the dilution, there are statistically rare
large spatial regions devoid of impurities and thus unaffected by the
disorder. The probability for finding such a region of volume $L^{d}$,
frequently referred to as an instanton, is
\begin{equation}
w\sim (1-p)^{\left( L/a_{0}\right) ^{d}}=\exp (-cL^{d})  \label{eq:prob}
\end{equation}%
with $c=-a_{0}^{-d}\ln (1-p)$. Below the clean critical point, the rare
regions can be locally in the ordered phase even though the bulk system is
not. At zero temperature, each rare region is equivalent to a
one-dimensional classical $O(N)$ model in a rod-like geometry: finite in the
$d$ space dimensions but infinite in imaginary time. For overdamped
dynamics, $z=2$, the interaction in imaginary time direction is of the form $%
(\tau -\tau ^{\prime })^{-2}$. One-dimensional continuous-symmetry $O(N)$
models with $1/\mathbf{x}^{2}$ interaction are known to be exactly \emph{at}
their lower critical dimension\cite{Joyce69,Dyson69, Bruno01}. Therefore, an
isolated rare region of linear size $L$ cannot independently undergo a phase
transition. Its energy gap depends exponentially on its volume (i.e., the
effective spin of the droplet)
\begin{equation}
\epsilon _{L}\sim \exp (-bL^{d})~.  \label{eq:gap}
\end{equation}%
Equivalently, the susceptibility of such a region diverges exponentially
with its volume. Combining (\ref{eq:prob}) and (\ref{eq:gap}) gives a
power-law density of states for the energy gap $\epsilon $ (to leading
exponential accuracy),
\begin{equation}
\rho (\epsilon )\propto \epsilon ^{c/b-1}=\epsilon ^{d/z^{\prime }-1}
\label{eq:dos}
\end{equation}%
where the second equality defines the customarily used dynamical exponent $%
z^{\prime }$ \cite{Young97}. It continuously varies with disorder strength
or distance from the clean critical point. Many results follow from this.
For instance, a region with a local energy gap $\epsilon $ has a local spin
susceptibility that decays exponentially in imaginary time, $\chi _{\text{loc%
}}(\tau \rightarrow \infty )\propto \exp (-\epsilon \tau )$. Averaging by
means of $\rho $ yields
\begin{equation}
\chi _{\text{loc}}^{\text{av}}(\tau \rightarrow \infty )\propto \tau
^{-d/z^{\prime }}.  \label{eq:chi_tau}
\end{equation}%
The temperature dependence of the static average susceptibility is then
\begin{equation}
\chi _{\text{loc}}^{\text{av}}(T)=\int_{0}^{1/T}d\tau \ \chi _{\text{loc}}^{%
\text{av}}(\tau )\propto T^{d/z^{\prime }-1}.  \label{eq:chi_T}
\end{equation}%
If $d<z^{\prime }$, the local zero-temperature susceptibility diverges, even
though the system is globally still in the disordered phase. Analogously,
the contribution of the rare regions to the specific heat $C$ can be
obtained from
\begin{equation}
\Delta E=\int d\epsilon ~\rho (\epsilon )~\epsilon ~e^{-\epsilon
/T}/(1+e^{-\epsilon /T})\propto T^{d/z^{\prime }+1}~  \label{eq:cv}
\end{equation}%
which gives $\Delta C\propto T^{d/z^{\prime }}$. Other observables can be
determined in a similar fashion. The power-law density of states (\ref%
{eq:dos}) in the Griffiths phase of a disordered itinerant $O(N)$ magnet and
the resulting quantum Griffiths singularities (\ref{eq:chi_tau}), (\ref%
{eq:chi_T}), (\ref{eq:cv}) are the central results of this Paper. They take
the same form as the quantum Griffiths singularities in undamped (clean $z=1$%
) random quantum Ising models \cite%
{McCoy69,dsf9295,YoungRieger96,Pich98,Motrunich00} and quantum Ising spin
glasses \cite{ThillHuse95,gbh96,RiegerYoung96}.

These scaling arguments suggest a general classification of Griffiths
phenomena in the vicinity of bulk phase transitions (at least those
described by Landau-Ginzburg-Wilson theories) with weak, random-$T_{c}$ or
random-mass type disorder. It is based on the effective dimensionality of
the rare regions. Three cases can be distinguished:

(i) If the rare regions are \emph{below} the lower critical dimensionality $%
d_{c}^{-}$ of the problem, their energy gap depends on their size via a
power law, $\epsilon _{L}\sim L^{-\psi }$. Since the probability for finding
a rare region is exponentially small in $L$, the low-energy density of
states in this first case is exponentially small. This leads to weak
\textquotedblleft classical\textquotedblright\ Griffiths singularities
characterized by an essential singularity in the free energy. This case is
realized in generic classical systems (where the rare regions are finite in
all directions and thus effectively zero-dimensional). It also occurs in
quantum rotor systems with Heisenberg symmetry and undamped ($z=1$) dynamics%
\cite{Read95,us_hsb04}. Here, the rare regions are equivalent to
one-dimensional classical Heisenberg models which are also below $%
d_{c}^{-}=2 $.

(ii) In the second class, the rare regions are exactly \emph{at} the lower
critical dimension. In this case, their energy gap shows an exponential
dependence [like (\ref{eq:gap})] on $L$. As shown above, this leads to a
power-law density of states and strong power-law quantum Griffiths
singularities. This second case is realized, e.g., in classical Ising models
with linear defects \cite{McCoy69,McCoyWu} and random quantum Ising models
(each rare regions corresponds to a one-dimensional classical Ising model)
\cite{dsf9295,YoungRieger96} as well as in the disordered itinerant quantum
Heisenberg magnets studied here (the rare regions are equivalent to
classical one-dimensional Heisenberg models with $1/\mathbf{x}^2$
interaction).

(iii) Finally, in the third class, the rare regions are \emph{above} the
lower critical dimension, i.e., they can undergo the phase transition
independently from the bulk system. In this case, the locally ordered rare
regions become truly static which leads to a smeared phase transition. This
happens, e.g., for classical Ising magnets with planar defects \cite%
{planar_ising} (the rare regions are effectively two-dimensional) or for
itinerant quantum Ising magnets \cite{smearing_prl,MillisMorrSchmalian}
where the rare regions are equivalent to classical one-dimensional Ising
models with $1/\mathbf{x}^2$ interaction.


\section{Rare regions in the large-$N$ limit}

\label{sec:large-N}

To complement the general scaling arguments and to obtain quantitative
estimates for the exponent $z^{\prime }$ we now perform an explicit
calculation of the Griffiths effects in the model (\ref{eq:action}) in the
large-$N$ limit. The approach is a generalization of Bray's treatment \cite%
{Bray87} of the classical case. In the large-$N$ limit, a clean system
undergoes a quantum phase transition for $g=g_{c}\propto \Lambda ^{2-d-z}$
with coupling constant $g=\frac{u}{\left\vert r_{0}\right\vert }$ and upper
momentum cut off $\Lambda $. For $g<g_{c}$, the clean system is in the
ordered state with the order parameter
\begin{equation}
\phi _{0,\mathrm{clean}}=[{N({g_{c}-g})/({g_{c}g})}]^{1/2},
\end{equation}%
and vanishing gap. In the random system, we consider a droplet of size $%
L^{d} $, devoid of impurities and determine its size dependent energy gap, $%
\epsilon $. It is determined by the equation of state $\epsilon \phi _{0}=h$%
, where
\begin{equation}
\epsilon =r_{0}+u\left\langle \phi ^{2}\right\rangle +\frac{u\phi _{0}^{2}}{N%
}.  \label{eq:eos}
\end{equation}%
$h$ is the field conjugate to the order parameter, $\phi _{0}=\left\langle
\phi \right\rangle $, of the droplet and
\begin{equation}
\left\langle \phi ^{2}\right\rangle =\sum_{\mathbf{q},\omega _{n}}\frac{%
TL^{-d}}{\epsilon +\mathbf{q}^{2}+\left\vert \omega _{n}\right\vert ^{2/z}}.
\label{eq:sum}
\end{equation}%
For $T>0$, both the $\mathbf{q}$ and $\omega $ sums are discrete.
Consequently, the order parameter $\phi _{0}=h/\epsilon $ vanishes for $%
h\rightarrow 0$ since $\epsilon \ $must remain positive to avoid a
divergence of the $\mathbf{q=0}$, $\omega _{n}=0$ contribution to $%
\left\langle \phi ^{2}\right\rangle $. Thus, classical droplets are below $%
d_{c}^{-}$. t $T=0$, a frequency integration must be performed and the $%
\epsilon \rightarrow 0$ limit becomes less singular. Yet, for $z<2$ droplets
remain below $d_{c}^{-}$ since the $\mathbf{q=0}$ contribution to $%
\left\langle \phi ^{2}\right\rangle $ still diverges as $\ L^{-d}\epsilon ^{%
\frac{z-2}{2}}$.\ For $z=2$ this term diverges only as $\ln \left( \epsilon
L^{2}\right) $ and, as expected, droplets with $z=2$ are marginal and
located at their lower critical dimension.

To quantify these arguments and to determine the dependence of $\epsilon $
on $L$ for $T=0$, we apply the finite size analysis of the large-$N$ theory
\cite{Brezin82} to the quantum limit. As shown in the appendix, we obtain
for $\varepsilon L^{2}\ll 1$ and $z=2$:
\begin{equation}
\left\langle \phi ^{2}\right\rangle =\frac{1}{g_{c}}-\frac{L^{-d}}{\pi }\ln
\left( \epsilon L^{2}\right) \,.  \label{eq:finites}
\end{equation}%
Inserting this into (\ref{eq:eos}) gives for small $\epsilon $:
\begin{equation}
\epsilon =L^{-2}\exp \left( -bL^{d}\right) ,  \label{eq:rL}
\end{equation}%
with $b=\pi \ \frac{g_{c}-g}{g_{c}g}=\frac{\pi }{N}\phi _{0,\mathrm{clean}%
}^{2}$. This explicitly verifies eq.\ (\ref{eq:gap}) in the large-$N$ limit.
For a given distance, b, to the clean critical point, only droplets larger
than a certain value contribute to quantum Griffiths behavior. For $z<2$,
the last term in (\ref{eq:finites}) is proportional to $L^{-d}\epsilon ^{%
\frac{z-2}{2}}$ and we obtain $\epsilon \propto L^{-\psi }$ with $\psi =%
\frac{2d}{2-z}$. For $z>2$, $\phi _{0}\neq 0$, i.e. even a finite droplet is
allowed to order at $T=0$. Since the onset of order depends on the size of
the droplet a smearing of the transition occurs. All this is in agreement
with our general expectation, discussed above.

\begin{figure}[tbp]
\label{FLOW} \includegraphics[width=8.5cm]{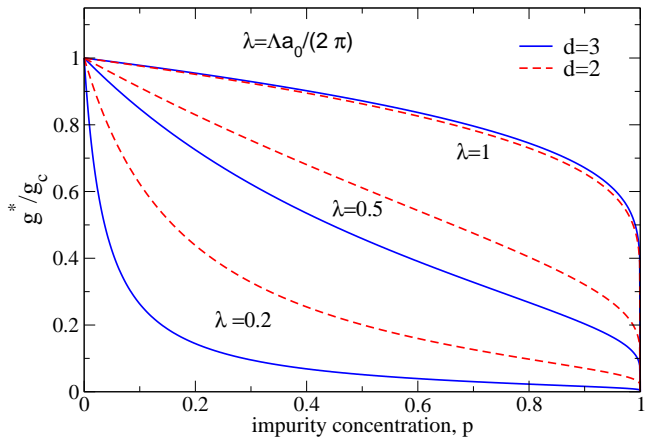}
\caption{Coupling constant $g^{\ast }/g_{c}$ below which quantum Griffiths
effects cause a diverging low energy density of states, as function of
disorder concentration, $p$, in two and three dimensions for various values
of $\protect\lambda =\frac{\Lambda a_{0}}{2\protect\pi }$.}
\end{figure}

Inserting our result for the coefficient $b$ into (\ref{eq:dos}), we obtain
an explicit expression for the Griffiths exponent
\begin{equation}
z^{\prime }=\frac{d\pi \ \left( g_{c}-g\right) a_{0}^{d}}{g_{c}g\ln
(1-p)^{-1}}.
\end{equation}%
The density of states (\ref{eq:dos}) diverges for $\epsilon \rightarrow 0$
if $z^{\prime }>d$. $z^{\prime }$ vanishes as one approaches the clean
critical point $g\rightarrow g_{c}$, but becomes larger as $\left(
g_{c}-g\right) /g$ grows. In Fig.\ 1, we plot the coupling constant $g^{\ast
}$, below which $z^{\prime }>d$, as function of the impurity concentration, $%
p$, for three different values of the non-universal number $a_{0}\Lambda $.
Quantum Griffiths effects dominate the low energy excitations for $g<g^{\ast
}$, provided that droplets are still sufficiently diluted and the system is
above the critical point, $g_{c}^{\mathrm{dis}}$, of the random system.
Observable quantum Griffiths effects exist unless $\Lambda a_{0}$ becomes
small.

At finite temperatures, a crossover occurs to weaker classical Griffiths
effects. To estimate the characteristic crossover temperature for $z=2$, we
decompose $\left\langle \phi ^{2}\right\rangle $, eq.\ (\ref{eq:sum}), into
its zero-temperature part and the more singular classical ($\omega _{n}=%
\mathbf{q}=0$) contribution:
\begin{equation}
\left\langle \phi ^{2}\right\rangle _{T}\simeq \left\langle \phi
^{2}\right\rangle _{T=0}+L^{-d}\frac{T}{\varepsilon _{L}}.  \label{eq:fT}
\end{equation}%
The crossover occurs when the classical term becomes comparable to $%
\left\langle \phi ^{2}\right\rangle _{T=0}$. We find that droplets with $L>$
$L_{0}\left( T\right) $, determined by $T=(b/\pi )L_{0}^{d-2}e^{-bL_{0}^{d}}$%
, behave classically and $\rho \left( \epsilon \right) $ is suppressed for $%
\epsilon <\epsilon _{0}=L_{0}^{-2}e^{-bL_{0}^{d}}$. Droplets smaller than $%
L_{0}\left( T\right) $ still follow the quantum dynamics. Quantum Griffiths
behavior persists as long as $\epsilon _{0}\left( T\right) <T$. This is
fulfilled for sufficiently low temperatures $T<T_{0}=f_{d}b^{2/d}$ with $%
f_{d}=\pi ^{-2/d}\exp (-\pi )$. Above the temperature $T_{0}$ the density of
states for $\omega >T$ is then given as
\begin{equation}
\rho _{class}\left( \varepsilon \right) \propto \exp \left( -\frac{A}{%
\varepsilon }\right)
\end{equation}%
with $A=\frac{\pi T}{ba_{0}^{d}}\ln (1-p)^{-1}$, in agreement with the
behavior for classical Heisenberg systems\cite{Bray87}. \ Instead of adding
the purely classical contribution to $\left\langle \phi ^{2}\right\rangle
_{T=0}$, the behavior at low $T$ may be described by explicitly calculating
the first low temperature corrections to $\left\langle \phi
^{2}\right\rangle _{T=0}$. We find \ that these low temperature corrections
behave as
\begin{equation}
\left\langle \phi ^{2}\right\rangle _{T}=\left\langle \phi ^{2}\right\rangle
_{T=0}+\frac{L^{-d}}{48\pi }\frac{T^{2}}{\ \left( 4\pi \varepsilon
_{L}\right) ^{2}}+...
\end{equation}%
An argumentation identical to the one given below Eq.\ref{eq:fT} yields that
these corrections become relevant above a temperature which behaves
identical to $T_{0}\propto b^{2/d}$, only with a different numerical
prefactor $f_{d}\simeq e^{-1}\left( 16\pi \sqrt{3}\right) ^{\frac{4}{d}}$.
The closer one one approaches the clean quantum phase transition, where $b$
vanishes, the narrower is the region of quantum Griffiths behavior. More
importanly, very close to the dirty critical point, $g_{c}^{\mathrm{dis}}$,
the crossover temperature is exponentially suppressed because the droplets
have a minimum size of the order of the bulk correlation length.


\section{Discussion and conclusions}

\label{sec:conclusions}

To summarize, we have studied quantum Griffiths effects in itinerant magnets
with continuous order parameter symmetry, using the itinerant
antiferromagnet as the primary example. We have shown that this system
displays strong power-law quantum Griffiths singularities. In this section
we will address a few open questions and important implications of the
results.

Our first point concerns the importance of Kondo physics. In eq.\ (\ref%
{eq:action}) we assumed, following Refs.\ \cite%
{Hertz76,Millis93,BelitzKirkpatrick96}, that spin degrees of freedom in
disordered metals can be described by quantum rotors with overdamped
dynamics. One might worry about the Kondo dynamics of the entire
dropletwhich is not included in the rotor approach (in Ref.\ \cite{Shah03}
it was shown that an extended magnetic structure in a metallic environment
can behave at low $T$ as an ansiotropic multi-channel Kondo problem). Our
theory is valid only if the $k$-channel Kondo behavior, presumably with
large $k\sim \left( L/a_{0}\right) ^{d}$, emerges only for $T_{K}\ll
\varepsilon \left( L\right) $. Using the results for the related problem of
a large droplet induced by a single magnetic impurity\cite{Larkin72,turlakov}
it indeed follows that the Kondo temperature $T_{K}\sim \varepsilon \left(
L\right) \exp \left( {-const.L^{d}}\right) $ is exponentially smaller than
the crossover scale for quantum Griffiths behavior and thus negligible.

We emphasize the difference between continuous spin symmetry and Ising
symmetry. For Ising symmetry, rare regions are \emph{above} the lower
critical dimension. They cease to tunnel and become static at sufficiently
low temperatures, leading to superparamagnetic behavior \cite%
{MillisMorrSchmalian} and, ultimately, to a smeared transition\cite%
{smearing_prl}. Quantum Griffiths behavior can at best occur in a transient
temperature window\cite{CastroNetoJones}. In contrast, for continuous
symmetry, the rare regions are \emph{exactly at} the lower critical
dimension and retain their dynamics, with a power-law low-energy density of
states. Quantum Griffiths effects dominate the low-temperature physics for $%
g^{\ast }>g>g_{c}^{\mathrm{dis}}$.

Griffiths phenomena in itinerant \emph{ferro}\-magnets require separate
attention because mode-coupling effects induce a long-range interaction of
the spin fluctuations \cite{kb_fm_dirty}. This can potentially change the
conditions for locally ordered droplets and thus the form of the Griffiths
effects.

Let us comment on measuring the predicted effects. Many of the heavy
electron systems displaying magnetic quantum phase transitions have a strong
spin anisotropy and are thus better described by Ising models. \textrm{Gd}%
-based intermetallics have a local Heisenberg symmetry, but the
hybridization between magnetic and conduction electrons is very small. This
yields a very low temperature, $T_{0}$, for the onset of quantum Griffiths
behavior. Most promising are $3d$-transition metal systems with weak
spin-orbit interaction and strong hybridization. A candidate is the $3d$
heavy fermion system LiV$_{2}$O$_{4}$\cite{LiV2O4A} where recent experiments
did show a broad distribution of relaxation rates \cite{LiV2O4B}. In
addition, the $XY$-version of our theory ($N=2$) directly predicts quantum
Griffiths behavior in disordered thin nanowires\cite{Rogachev03} close to
the metal-superconductor quantum phase transition\cite{Sachdev04}, with
direct impact on the conductance fluctuations of these systems. Our results
hopefully motivate further the search for new and unconventional behavior in
dissipative quantum systems with continuous order parameter symmetry.

This letter has focused on the Griffiths region above the dirty critical
point. There is, however, a possible connection to the properties of the
quantum critical point itself. It is known that the quantum critical points
of undamped random quantum Ising models, which also display power-law
quantum Griffiths effects, are of exotic infinite-randomness type \cite%
{McCoy69,dsf9295,YoungRieger96,Pich98,Motrunich00}. The underlying
strong-disorder renormalization group \cite{dsf9295,SDRG} supports a close
connection between the quantum Griffiths effects and the exotic critical
properties. This suggests that the quantum critical point of disordered
itinerant Heisenberg magnets may also be of infinite-randomness type.

We acknowledge helpful discussions with D. Belitz, A. Castro-Neto, T.R.
Kirkpatrick, A.J. Millis, D.K. Morr and R. Sknepnek. This work was supported
in part by the NSF under grant No. DMR-0339147 (T.V.) and by Ames
Laboratory, operated for the U.S. Department of Energy by Iowa State
University under Contract No. W-7405-Eng-82 (J.S.).

\section{Finite size analysis of the large $N$ theory}

In this appendix we summarize the derivation of Eq.\ref{eq:finites} by
generalizing the finite size analysis of Ref.\cite{Brezin82} to the quantum
case. The order parameter fluctuations
\begin{equation}
\left\langle \phi ^{2}\right\rangle =\sum_{\mathbf{q},\omega _{n}}\frac{%
TL^{-d}}{\epsilon +\mathbf{q}^{2}+\left\vert \omega _{n}\right\vert ^{2/z}}
\end{equation}%
determine the value of \ $\epsilon \left( L\right) $ through $\epsilon
=r_{0}+u\left\langle \phi ^{2}\right\rangle +\frac{u\phi _{0}^{2}}{N}$. \ We
consider a droplet of size $L^{d}$ and made the assumption that there are no
impurities inside of it. Then we are allowed to decsribe this droplet as a
clean, but finite system. Considering the zero temperature limit, the
frequency summation becomes an integration. The discrete sum over momenta is
then analyzed using the Poisson summation formula and we obtain $%
\left\langle \phi ^{2}\right\rangle =\left\langle \phi ^{2}\right\rangle
_{\infty }+J$ with%
\begin{equation}
\ \left\langle \phi ^{2}\right\rangle _{\infty }=\int \frac{d^{D}q}{\left(
2\pi \right) ^{d}}\int \frac{d\omega }{2\pi }\frac{1}{q^{2}+\left\vert
\omega \right\vert ^{2/z}+\varepsilon }
\end{equation}%
as well as
\begin{equation}
J=\sum_{\mathbf{n\neq 0}}S\left( \mathbf{n}\right)  \label{eq:J(L)}
\end{equation}%
with%
\begin{equation}
S\left( \mathbf{n}\right) =\int \frac{d^{d}q}{\left( 2\pi \right) ^{d}}\int
\frac{d\omega }{2\pi }\frac{e^{i\mathbf{q\cdot n}L}}{q^{2}+\left\vert \omega
\right\vert ^{2/z}+\varepsilon }.
\end{equation}%
Here $\mathbf{n=}\left( n_{1},...n_{d}\right) $ is a $d$-component vector
integer components.

As long as $d+z>2$, which we assume throughout this paper, it holds for
small $\varepsilon $ that $\left\langle \phi ^{2}\right\rangle _{\infty
}\simeq $ $\frac{1}{g_{c}}$, where $g_{c}$ is the critical coupling constant
of the clean bulk quantum phase transition, i.e. for $L\rightarrow \infty $.
The corrections behave as $\varepsilon ^{\frac{d+z-2}{2}}$ for $d+z<4$, $\ $%
as $\varepsilon \Lambda ^{d+z-4}$ for $d+z>4$, and \ as $\varepsilon \log
\left( \Lambda ^{2}/\varepsilon \right) $ for $d+z=4$, i.e. vanish as $%
\varepsilon \rightarrow 0$.

Next we analyze the sum over $\mathbf{n}$ in Eq.\ref{eq:J(L)} which takes
into account the finite size of the droplet. Using $\frac{1}{x}%
=\int_{0}^{\infty }d\alpha \exp \left( -\alpha x\right) $, it follows%
\begin{eqnarray*}
S\left( \mathbf{n}\right)  &=&\int_{0}^{\infty }d\alpha e^{-\alpha
\varepsilon }\int \frac{d\omega d^{d}q}{\left( 2\pi \right) ^{d+1}}e^{i%
\mathbf{q\cdot n}L}e^{-\alpha \left( q^{2}+\left\vert \omega \right\vert
^{2/z}\right) } \\
&=&\frac{\Gamma \left( 1+\frac{z}{2}\right) }{2^{d}\pi ^{\frac{d+2}{2}}}%
\int_{0}^{\infty }\frac{d\alpha e^{-\alpha \varepsilon }}{\alpha ^{\left(
d+z\right) /2}}\ \prod_{\mu =1,..,d}e^{-\frac{n_{\mu }^{2}L^{2}}{%
4\alpha }}
\end{eqnarray*}%
where we performed the momentum and frequency integrations. Substituting $%
t=4\pi \alpha L^{-2}$ it we obtain for $J$:%
\begin{equation}
J=\frac{\Gamma \left( 1+\frac{z}{2}\right) L^{2-\left( d+z\right) }}{%
2^{2-z}\pi ^{\frac{4-z}{2}}}A\left( \varepsilon L^{2}\right) \
\end{equation}%
with
\begin{equation}
A\left( \rho \right) =\int_{0}^{\infty }\frac{dt}{t^{\left( d+z\right) /2}}%
e^{-\frac{\rho t}{4\pi }}\left( B\left( \frac{1}{t}\right) ^{d}-1\right)
\end{equation}%
and
\begin{equation}
B\left( t\right) =\sum_{n=-\infty }^{\infty }\exp \left( -\pi tn^{2}\right)
\end{equation}%
It holds $B\left( t\right) =\left( \frac{1}{t}\right) ^{1/2}B\left( \frac{1}{%
t}\right) $ which leads to
\begin{eqnarray}
A\left( \rho \right)  &=&\int_{1}^{\infty }\frac{dt}{t^{z/2}}e^{-\frac{\rho t%
}{4\pi }}\left( 1-t^{-d/2}\right)   \nonumber \\
&&+\int_{1}^{\infty }dt\left( B\left( t\right) ^{d}-1\right)
\label{eq:A(rho)} \\
&&\times \left[ t^{\left( d+z-4\right) /2}e^{-\frac{\rho }{4\pi t}%
}+t^{-z/2}e^{-\frac{\rho t}{4\pi }}\right]   \nonumber
\end{eqnarray}%
For $z\leq 2$, the dominant contribution to $A\left( \rho \right) $ comes
from the first integral in Eq.\ref{eq:A(rho)} which can be evaluated
explicitly. The second integral is finite as $\rho =\varepsilon
L^{2}\rightarrow 0$ for all $z$. For small $\rho $ and $z<2$ follows for the
leading contributions%
\begin{equation}
A\left( \rho \right) =\frac{\Gamma \left( 1-\frac{z}{2}\right) }{\left( 4\pi
\right) ^{\frac{z}{2}-1}}\rho ^{-\frac{2-z}{2}}+\mathcal{O}\left( \rho
^{0}\right)
\end{equation}%
while
\begin{equation}
A\left( \rho \right) =-\log \left( \frac{\rho }{4\pi }\right) +\mathcal{O}%
\left( \rho ^{0}\right)
\end{equation}%
if $z=2$. If $z>2$, $A\left( \rho \rightarrow 0\right) $ remains finite.

Collecting the various contributions to $\left\langle \phi ^{2}\right\rangle
$ yields for $z=2$:
\begin{equation}
\left\langle \phi ^{2}\right\rangle =\frac{1}{g_{c}}-\frac{\ L^{-d}}{\pi }%
\log \left( \varepsilon L^{2}\right)
\end{equation}%
which is the result Eq.\ref{eq:finites} given above.

\end{document}